\documentclass[twocolumn]{aastex701}

\shorttitle{A High-Likelihood Polar Interstellar Meteor Candidate}
\shortauthors{Cloete \& Loeb}

\newcommand{\vinfsun}{v_{\infty,\odot}}
\newcommand{\vinfearth}{v_{\infty,\oplus}}
\newcommand{\vescsun}{v_{\mathrm{esc},\odot}}
\newcommand{\vhel}{v_\odot}
\newcommand{\pbound}{p_\mathrm{bound}}

\newcommand{\kms}{~\mathrm{km\,s^{-1}}}
\newcommand{\okina}{\textquoteleft}
\newcommand{\kmssq}{~\mathrm{km^2\,s^{-2}}}
\newcommand{\polarim}{Polar-IM}

\graphicspath{{../../figures/}}

\begin{document}

\title{A Polar Interstellar Meteor Candidate from 04-01-2026 in CNEOS}

\author{Richard Cloete}
\affiliation{Astronomy Department, Harvard University, 60 Garden Street, Cambridge, MA, 02138, USA}
\email[]{richardcloete@cfa.harvard.edu}

\author{Abraham Loeb}
\affiliation{Astronomy Department, Harvard University, 60 Garden Street, Cambridge, MA, 02138, USA}
\email[show]{aloeb@cfa.harvard.edu}

\begin{abstract}
We report a newly identified polar interstellar meteor candidate, labeled \polarim{}, detected on 2026-04-01 02:13:14~UTC at latitude $-41.9\degr$, longitude $-54.7\degr$, and altitude 90.5~km over the South Atlantic ocean, east of Argentina. We transform the reported Earth-fixed velocity vector $(+3.6,\,-34.6,\,+59.8)\kms$ to an inertial geocentric state, remove Earth's gravitational acceleration with a two-body hyperbolic model, add the JPL Horizons heliocentric velocity of Earth, and test the resulting heliocentric orbit against solar escape speed. The final velocity component in the polar ($z$) direction of $+47.09\kms$ exceeds by itself the local solar escape speed $\vescsun = 42.14\kms$. The full heliocentric speed is $\vhel = 51.73\kms$, corresponding to positive heliocentric specific energy $\varepsilon_\odot = +450.1\kmssq$, heliocentric excess speed $\vinfsun = 30.00\kms$, and a two-body inclination $i=89.4\degr$. We propagate measurement uncertainty through 1{,}000{,}000 Monte-Carlo realizations using the empirical post-2018 low-discrepancy CNEOS error model of \citet{PenaAsensio2025}, with $\sigma_v = 0.55\kms$, $\sigma_\mathrm{RA} = 1.35\degr$, and $\sigma_\mathrm{Dec} = 0.84\degr$. No realization yields a bound heliocentric orbit, giving a statistical confidence on the interstellar fraction of $>99.9997\%$. The Monte-Carlo margin above escape is $\langle\Delta\rangle = 9.60 \pm 0.75\kms$, corresponding to a $12.82\sigma$ margin-to-scatter ratio under the adopted perturbation model. The result identifies \polarim{} as the highest-margin post-2018 candidate in the CNEOS catalog.
\end{abstract}

\keywords{interstellar objects --- meteors --- meteoroids --- fireballs}

\section{Introduction}
\label{sec:intro}

The discoveries of 1I/\okina Oumuamua \citep{Meech2017}, 2I/Borisov \citep{Guzik2020}, and 3I/ATLAS \citep{seligman2025discovery} established that macroscopic bodies from other stellar systems pass through the inner Solar System. Telescopic surveys are sensitive to the larger members of this population, but smaller interstellar debris should be far more numerous \citep[e.g.,][]{Do2018,jewitt2023interstellar} and may be detectable only when it enters Earth's atmosphere as a bright fireball. A fireball produced by an interstellar meteoroid would provide a direct measurement of the object's encounter velocity, and in rare cases may leave recoverable material that can be studied in the laboratory \citep{Loeb2024}.

Identifying such events is challenging. A meteoroid is interstellar only if its pre-encounter heliocentric orbit is gravitationally unbound from the Sun, which requires a velocity measurement precise enough to distinguish a true excess above solar escape speed from a bound orbit shifted by measurement error. The NASA Center for Near-Earth Object Studies (CNEOS) fireball database is uniquely valuable because it reports global bolide detections from U.S.\ Government sensors, including three-component velocity vectors for a subset of events. However, CNEOS does not publish per-event uncertainty estimates or covariance matrices. Any claim of interstellar origin must therefore be conditional on an external uncertainty model and must propagate that uncertainty through the full coordinate and energy calculation.

Earlier interstellar meteor claims, including the proposed candidates from 2014 \citep{Siraj1} and 2017 \citep{SirajLoeb2022,Pena22}, demonstrated both the promise and the controversy of this approach. The central concern has not been the orbital-energy criterion itself, but the reliability of the reported velocity vectors, especially for pre-2018 events and nominally hyperbolic entries \citep{Devillepoix2019,SocasNavarro2024}. Recent empirical calibration work by \citet{PenaAsensio2025}, based on cross-matches between CNEOS detections and independent ground-based fireball networks, provides a practical way forward. Their analysis separates the catalog into a high-discrepancy regime, where speed and radiant errors can be large, and a low-discrepancy regime, dominated by post-2018 events, where the reported velocities are consistent with much smaller uncertainties. The same calibration work nevertheless cautions that hyperbolic CNEOS candidates should be interpreted carefully because apparent hyperbolic signatures can reflect measurement errors.

This paper applies that calibrated framework to a single CNEOS event: the fireball detected on 2026-04-01 02:13:14~UTC, which we designate \polarim{}. The event is a post-2018 low-discrepancy case with a very large reported geocentric speed, a strongly positive velocity component along the terrestrial rotation axis, and a nominal heliocentric speed well above solar escape. It was flagged by one of the authors (A.L) in the CNEOS sample, following on the earlier post-2018 screening that reported the 2022-07-28 and 2025-02-12 candidates \citep{CloeteLoeb2026}.

Our analysis transforms the CNEOS Earth-fixed velocity vector to an inertial geocentric frame, removes Earth's gravitational influence to estimate the geocentric hyperbolic excess velocity, adds the JPL Horizons heliocentric velocity of Earth, and evaluates the resulting heliocentric specific orbital energy. Measurement uncertainty is then propagated by perturbing the inertial geocentric speed and radiant direction under the adopted low-discrepancy error model.

Our principal result is that \polarim{} remains unbound in all 1{,}000{,}000 Monte-Carlo realizations. Its mean Monte-Carlo heliocentric speed exceeds the solar escape speed at Earth's heliocentric distance by $9.60 \pm 0.75\kms$, corresponding to a model-dependent $12.82\sigma$ margin-to-scatter ratio. The finite-sample 95\% Monte-Carlo upper limit on the bound probability is $\pbound < 3.0 \times 10^{-6}$, equivalent to a lower confidence bound of $>99.9997\%$ on the unbound fraction, conditional on the adopted Gaussian low-discrepancy uncertainty model. These numbers make \polarim{} a high-likelihood interstellar meteor candidate.

This paper is organized as follows. Section~\ref{sec:data} describes the CNEOS and JPL Horizons data sources used in the analysis. Section~\ref{sec:methods} summarizes the coordinate transformations, gravitational correction, heliocentric energy test, and Monte-Carlo uncertainty propagation. Section~\ref{sec:results} reports the deterministic and Monte-Carlo results for \polarim{}. Section~\ref{sec:discussion} discusses systematics, limitations, and follow-up priorities. Section~\ref{sec:summary} summarizes the conclusions.

\section{Data Sources}
\label{sec:data}

\subsection{CNEOS Fireball Database}

The primary observational input is the NASA/JPL CNEOS fireball database, accessed through the SSD Fireball API \citep{CNEOSFireballAPI}. We consider events with geographic location, altitude, and three-component velocity information, using the API fields corresponding to event time, latitude, longitude, altitude, radiated energy, impact energy, and velocity components. The API field table labels $v_x$, $v_y$, and $v_z$ as Earth-centered entry-velocity components, while the public CNEOS fireball-table description defines the velocity-component frame as geocentric Earth-fixed, with the $z$-axis directed along Earth's rotation axis toward the celestial north pole, the $x$-axis in the equatorial plane toward the prime meridian, and the $y$-axis completing the right-handed coordinate system \citep{CNEOSFireballs}. We therefore treat the reported components as Earth-fixed Cartesian components, denoted throughout as $(v_x, v_y, v_z)_\mathrm{ECEF}$.

Only events with the full set of quantities required for orbit determination are processed. The analyzed CNEOS sample contains 356 events with complete location, altitude, and velocity-component data. Seven of those events have positive nominal heliocentric specific energy under the deterministic classifier. This study does not attempt to adjudicate all seven candidates; it focuses on \polarim{}, the event detected on 2026-04-01 02:13:14~UTC. The CNEOS metadata for this event are listed in the upper section of Table~\ref{tab:inputs}.

\begin{figure}
\centering
\includegraphics[width=\columnwidth]{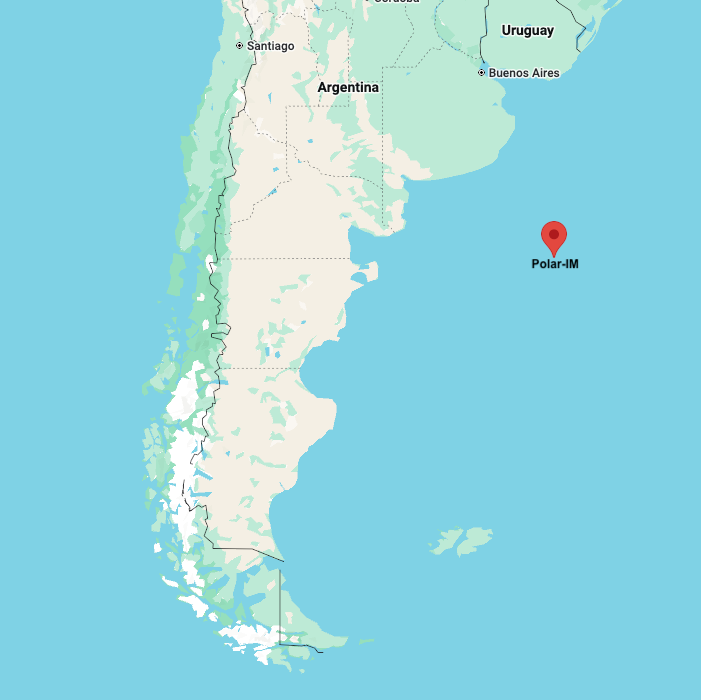}
\caption{Reported geographic location of \polarim{}, the 2026-04-01 CNEOS fireball. The marker shows the event position at latitude $-41.9\degr$ and longitude $-54.7\degr$, over the South Atlantic east of Argentina.}
\label{fig:location}
\end{figure}

The coordinates place the event over the South Atlantic east of Argentina (Figure~\ref{fig:location}). The impact energy is below the secondary high-energy threshold used by the adopted uncertainty model, but the event qualifies for the low-discrepancy regime by epoch because it occurred after 2018. CNEOS does not provide event-specific velocity uncertainties, radiant uncertainties, or a covariance matrix. The uncertainty propagation therefore uses the externally calibrated low-discrepancy error model summarized in Section~\ref{sec:errormodel}.

\subsection{JPL Horizons Earth Ephemeris}

To convert a geocentric incoming velocity into a heliocentric velocity, the analysis requires Earth's heliocentric state at the event epoch. This state is obtained from JPL Horizons \citep{Giorgini1996} using the Astroquery \texttt{jplhorizons} interface \citep{Ginsburg2019}. For each fireball event, we query the heliocentric vector state of Earth at the event time, with the Sun as the origin and the J2000 equatorial frame as the reference frame. The Horizons-derived Earth state used for \polarim{} is given in the lower section of Table~\ref{tab:inputs}.

The ephemeris uncertainty is negligible for this analysis. The dominant uncertainty is the CNEOS velocity measurement, which is at the km\,s$^{-1}$ level under the empirical calibration used here.

\subsection{Reproducibility}

The calculation uses public event parameters from the CNEOS fireball database and the public JPL Horizons ephemeris service. The deterministic part of the analysis can be reproduced from the CNEOS event time, location, altitude, reported velocity components, and impact energy listed in Table~\ref{tab:inputs}, together with the coordinate transformations and two-body energy tests described below. The Monte-Carlo interpretation additionally requires adopting the low-discrepancy uncertainty model summarized in Section~\ref{sec:errormodel}.

\begin{deluxetable}{lc}
\tablecaption{Adopted inputs for \polarim{}: CNEOS metadata, the JPL Horizons heliocentric Earth state at the event epoch (J2000 equatorial), and the adopted low-discrepancy uncertainty model.\label{tab:inputs}}
\tablehead{\colhead{Quantity} & \colhead{Value}}
\startdata
\cutinhead{CNEOS metadata}
Date/time (UTC) & 2026-04-01 02:13:14 \\
Latitude & $-41.9\degr$ \\
Longitude & $-54.7\degr$ \\
Altitude (km) & 90.5 \\
Radiated energy (J) & $2.4\times10^{10}$ \\
API \texttt{energy} value ($10^{10}$ J) & 2.4 \\
Impact energy (kt) & 0.086 \\
$(v_x, v_y, v_z)_\mathrm{ECEF}$ (km\,s$^{-1}$) & $(+3.6,\,-34.6,\,+59.8)$ \\
Uncertainty regime & Low-discrepancy \\
\cutinhead{JPL Horizons Earth state}
$x$ (km) & $-146{,}717{,}131.36$ \\
$y$ (km) & $-26{,}177{,}741.35$ \\
$z$ (km) & $-11{,}346{,}517.33$ \\
$v_x$ (km\,s$^{-1}$) & $+5.203$ \\
$v_y$ (km\,s$^{-1}$) & $-26.920$ \\
$v_z$ (km\,s$^{-1}$) & $-11.668$ \\
\cutinhead{Low-discrepancy error model}
$\sigma_v$ (km\,s$^{-1}$) & 0.55 \\
$\sigma_\mathrm{RA}$ & $1.35\degr$ \\
$\sigma_\mathrm{Dec}$ & $0.84\degr$ \\
$\sigma_{\theta,\mathrm{axis}}$ & $1.124\degr$ \\
\enddata
\end{deluxetable}

\section{Methods}
\label{sec:methods}

\subsection{Analysis Workflow}

The objective of the analysis workflow is to determine whether a CNEOS fireball's pre-encounter heliocentric trajectory was gravitationally bound or unbound with respect to the Sun. For each event with complete metadata, the calculation proceeds through seven stages: (1)~ingest the CNEOS event time, geographic coordinates, altitude, impact energy, and Earth-fixed velocity components; (2)~convert the reported geodetic location and Earth-fixed velocity into an inertial geocentric state in the Geocentric Celestial Reference System (GCRS); (3)~remove Earth's gravitational acceleration using a two-body hyperbolic model to estimate the geocentric hyperbolic excess velocity $\mathbf{v}_{\infty,\oplus}$; (4)~retrieve Earth's heliocentric velocity from JPL Horizons at the event epoch; (5)~add Earth's heliocentric velocity to the geocentric excess velocity to obtain the impactor's heliocentric velocity $\mathbf{v}_\odot$; (6)~compute the heliocentric specific orbital energy, solar escape speed, and heliocentric excess speed; and (7)~propagate CNEOS velocity uncertainty through the full calculation using a Monte-Carlo speed-plus-radiant perturbation model. Steps 1--6 produce the deterministic classification. Step 7 quantifies how robust that classification is under the adopted empirical uncertainty model.

\subsection{Earth-Fixed to Inertial Geocentric Coordinates}

The CNEOS velocity components are interpreted as Earth-fixed Cartesian components, following the public CNEOS fireball-table frame definition \citep{CNEOSFireballs} and the Earth-fixed convention used by \citet{PenaAsensio2025}. For each event, we construct an Astropy \citep{Astropy2022} \texttt{EarthLocation} from the reported latitude, longitude, and altitude. The geodetic location is evaluated at the event UTC time and represented as an ITRS position. The CNEOS velocity vector is attached to this ITRS position as a Cartesian differential, producing a six-dimensional Earth-fixed state $(\mathbf{r}_\mathrm{ITRS}, \mathbf{v}_\mathrm{ITRS})$. Astropy is then used to transform this state into the GCRS frame at the same epoch, $(\mathbf{r}_\oplus, \mathbf{v}_\oplus)_\mathrm{GCRS}$.

This transformation accounts for Earth rotation and the standard precession, nutation, and polar-motion corrections available through Astropy's Earth-orientation machinery. The analysis uses the bundled IERS Earth-orientation tables available to Astropy; any resulting Earth-orientation approximation error is far below the CNEOS velocity uncertainties relevant here.

\subsection{Removing Earth's Gravity}

The velocity measured near peak brightness includes acceleration acquired in Earth's gravitational well. To estimate the incoming velocity before Earth encounter, we account for Earth's gravity with a two-body hyperbolic approximation. The Earth-centered specific orbital energy is
\begin{equation}
\varepsilon_\oplus = \frac{1}{2}|\mathbf{v}_\oplus|^2 - \frac{\mu_\oplus}{|\mathbf{r}_\oplus|},
\end{equation}
where $\mu_\oplus = 398{,}600.44\ \mathrm{km^3\,s^{-2}}$. For a hyperbolic Earth encounter, $\varepsilon_\oplus > 0$, and the geocentric excess speed is $\vinfearth = \sqrt{2\varepsilon_\oplus}$.

The direction of $\mathbf{v}_{\infty,\oplus}$ is obtained from the hyperbolic asymptote geometry. We compute the angular-momentum vector $\mathbf{h} = \mathbf{r}_\oplus \times \mathbf{v}_\oplus$ and the eccentricity vector
\begin{equation}
\mathbf{e} = \frac{\mathbf{v}_\oplus \times \mathbf{h}}{\mu_\oplus} - \frac{\mathbf{r}_\oplus}{|\mathbf{r}_\oplus|}.
\end{equation}
For $e = |\mathbf{e}| > 1$, the incoming asymptote is selected as the branch most aligned with the observed inertial geocentric velocity, giving $\mathbf{v}_{\infty,\oplus} = \vinfearth\,\hat{\mathbf{s}}_\mathrm{in}$. For \polarim{}, this step gives a nominal inertial geocentric speed of 69.10~km\,s$^{-1}$ and a geocentric excess speed of 68.20~km\,s$^{-1}$.

\subsection{Heliocentric Energy Test}

The impactor's heliocentric velocity at Earth's orbital distance is computed by vector addition, $\mathbf{v}_\odot = \mathbf{v}_{\infty,\oplus} + \mathbf{v}_E$, where $\mathbf{v}_E$ is Earth's heliocentric velocity from JPL Horizons. The heliocentric specific orbital energy is
\begin{equation}
\varepsilon_\odot = \frac{1}{2}|\mathbf{v}_\odot|^2 - \frac{\mu_\odot}{r_E},
\end{equation}
where $r_E$ is Earth's heliocentric distance at the event epoch and $\mu_\odot$ is the solar gravitational parameter. Equivalently, the event is unbound when
\begin{equation}
|\mathbf{v}_\odot| > \vescsun(r_E) = \sqrt{\frac{2\mu_\odot}{r_E}}.
\end{equation}
For unbound events, the heliocentric excess speed is $\vinfsun = \sqrt{|\mathbf{v}_\odot|^2 - \vescsun^2}$, the speed the object would retain at infinity in the two-body solar approximation.

\subsection{Adopted CNEOS Uncertainty Model}
\label{sec:errormodel}

CNEOS does not publish per-event velocity covariances, so the statistical interpretation depends on an external error model. We adopt the empirical calibration of \citet{PenaAsensio2025}, which compares CNEOS fireballs against independently measured ground-based fireball-network trajectories. The calibration identifies a low-discrepancy regime associated primarily with post-2018 events, plus some high-energy earlier events. We classify an event as low-discrepancy if it occurred in 2018 or later, or if its reported impact energy is at least 0.45~kt. \polarim{} is therefore low-discrepancy by date. Because the calibration sample predates this 2026 event, the statistical interpretation assumes that the post-2018 CNEOS sensor and processing performance remained in the same low-discrepancy regime through the event epoch.

For low-discrepancy events, the adopted one-sigma uncertainties are $\sigma_v = 0.55\kms$ (speed magnitude), $\sigma_\mathrm{RA} = 1.35\degr$, and $\sigma_\mathrm{Dec} = 0.84\degr$ (radiant direction). The angular perturbation is implemented in the tangent plane to the velocity direction with an effective per-axis uncertainty
\begin{equation}
\sigma_{\theta,\mathrm{axis}} = \sqrt{\frac{\sigma_\mathrm{RA}^2 + \sigma_\mathrm{Dec}^2}{2}} \approx 1.12\degr.
\end{equation}
This isotropic tangent-plane approximation intentionally treats the right-ascension and declination uncertainties as angular sky-plane errors without applying an additional $\cos\delta$ factor to the right-ascension term. For \polarim{}, the nominal geocentric radiant has $\mathrm{Dec}\approx -59.5\degr$, so multiplying the right-ascension term by $\cos|\delta|$ would reduce the effective angular perturbation. The adopted isotropic value is therefore conservative for this event.

All probability statements in this study are conditional on this low-discrepancy Gaussian perturbation model. The model tests robustness to small speed and radiant perturbations drawn from the calibration; it does not by itself test rare event-specific, non-Gaussian failures of the reported velocity vector.

\subsection{Monte-Carlo Propagation}
\label{sec:mcprop}

The Monte-Carlo simulation perturbs the inertial geocentric velocity $\mathbf{v}_\oplus$, not the raw Earth-fixed components. Each realization proceeds as follows: (1)~draw a perturbed speed $v' = |\mathbf{v}_\oplus| + \mathcal{N}(0, \sigma_v)$; (2)~draw a two-dimensional radiant perturbation on the tangent plane perpendicular to the nominal velocity direction, using $\sigma_{\theta,\mathrm{axis}}$; (3)~construct the perturbed inertial velocity vector $\mathbf{v}'_\oplus$; (4)~re-run the Earth-gravity removal and heliocentric energy test; and (5)~record whether the perturbed realization is bound or unbound with respect to the Sun.

For \polarim{}, we use $N = 1{,}000{,}000$ realizations and NumPy \citep{numpy2020} random seed 1. The Monte-Carlo analysis records $k_\mathrm{bound}$, the number of realizations with $\varepsilon_\odot \leq 0$; $\hat{p}_\mathrm{iso} = 1 - k_\mathrm{bound}/N$, the observed unbound fraction; the $\pbound$ upper limit, which for $k_\mathrm{bound} = 0$ follows from the rule of three, $\pbound \lesssim 3/N$ at 95\% confidence; the exact Clopper-Pearson upper limit, $1-(0.05)^{1/N}$, for the zero-bound case; the equivalent lower confidence bound on the unbound fraction, $1-\pbound$; the Monte-Carlo heliocentric speed distribution $\langle\vhel\rangle \pm \sigma_{\vhel}$; the Monte-Carlo heliocentric excess-speed distribution $\langle\vinfsun\rangle \pm \sigma_{\vinfsun}$; the speed margin above solar escape $\Delta = \vhel - \vescsun$; and the margin-to-scatter ratio $z_\Delta = \langle\Delta\rangle / \sigma_\Delta$ under the adopted perturbation model. The numerical Monte-Carlo outcomes and finite-sample confidence limits are reported in Section~\ref{sec:results}. Those confidence limits are model-conditional sampling bounds, not direct physical probabilities independent of the adopted error model.

\subsection{Declared Assumptions}

The analysis makes the following modeling assumptions: (1)~the CNEOS velocity components are interpreted as Earth-fixed Cartesian components following the public CNEOS fireball-table frame definition; (2)~the measured velocity is taken at or near peak brightness and is not corrected for atmospheric deceleration, which is conservative for interstellar classification because deceleration lowers the measured speed relative to the pre-atmospheric speed; (3)~Earth's gravity is accounted for with a two-body model, neglecting lunar and higher-order geopotential perturbations; (4)~solar binding is evaluated with a two-body solar escape-speed criterion at Earth's heliocentric distance; (5)~JPL Horizons ephemeris errors are negligible relative to CNEOS velocity errors; and (6)~the Monte-Carlo uncertainties are Gaussian and based on the calibrated low-discrepancy CNEOS error model; catastrophic, event-specific vector errors are not included in the Monte-Carlo model. These assumptions are revisited in Section~\ref{sec:discussion}.

\section{Results}
\label{sec:results}

\subsection{Catalog Context}

\begin{figure}
\centering
\includegraphics[width=\columnwidth]{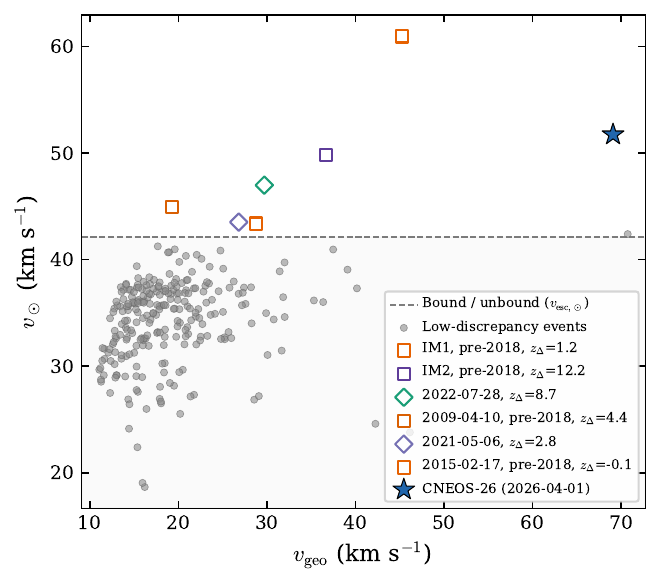}
\caption{Heliocentric speed $\vhel$ versus geocentric speed $v_\mathrm{geo}$ for CNEOS fireballs with complete velocity vectors. The horizontal dashed line marks the bound/unbound boundary at the solar escape speed $\vescsun \approx 42\kms$. Grey circles are low-discrepancy events with bound nominal orbits; nominal candidates are highlighted individually. \polarim{} lies well above the boundary and is the highest-margin post-2018 event in the analyzed sample ($z_\Delta = 12.82$; model-conditional $\pbound < 3.0\times10^{-6}$).}
\label{fig:scatter}
\end{figure}

The analyzed CNEOS sample contains 356 fireballs with complete event location, altitude, and velocity-component information. Seven events have positive nominal heliocentric specific energy and are therefore classified as candidate interstellar meteors by the deterministic energy test. Five of these seven are in the adopted low-discrepancy regime. \polarim{} is the focus of this paper because it is a new post-2018 low-discrepancy candidate, has a nearly polar two-body heliocentric orbit, and has the largest margin above solar escape among the post-2018 candidates in the analyzed sample. One nominal candidate, the 2015-02-17 event, has a positive deterministic margin but a slightly negative Monte-Carlo mean margin under the high-discrepancy error model; this distinction explains why nominal-candidate and Monte-Carlo-margin rankings need not be identical for borderline pre-2018 events. Figure~\ref{fig:scatter} places the event in the context of the full catalog.

\subsection{Deterministic Kinematic Results}

The reported velocity vector has a very large magnitude and a dominant positive $z$ component, $(v_z)_\mathrm{ECEF}=+59.8\kms$, in the CNEOS Earth-fixed convention. After transformation to the inertial geocentric frame, the nominal geocentric speed is 69.10~km\,s$^{-1}$ and the geocentric excess speed is 68.20~km\,s$^{-1}$. Adding Earth's heliocentric velocity yields a heliocentric velocity vector $(v_x, v_y, v_z)_\odot = (-20.97,\,-4.28,\,+47.09)\kms$ in the J2000 equatorial frame and a heliocentric speed $\vhel = 51.73\kms$, compared with a solar escape speed of $\vescsun = 42.14\kms$ at Earth's heliocentric distance. The single positive heliocentric $z$ component is already larger than the scalar local escape speed; therefore, under the adopted velocity-frame convention, the full three-dimensional heliocentric velocity vector is necessarily unbound in the deterministic calculation. The nominal speed exceeds solar escape by $\Delta = 9.59\kms$, the heliocentric specific energy is strongly positive, $\varepsilon_\odot = +450.1\kmssq$, and the heliocentric excess speed is $\vinfsun = 30.00\kms$. These quantities are collected in Table~\ref{tab:summary}.

The classification is driven by the vector direction as well as by the high reported geocentric speed. The geocentric excess speed of 68.20~km\,s$^{-1}$ is below the approximate maximum atmospheric-entry speed of a Sun-bound meteoroid after Earth gravitational focusing, about 72--73~km\,s$^{-1}$. \polarim{} is therefore not classified as interstellar by a simple geocentric speed ceiling. Instead, the inferred incoming direction places most of the heliocentric velocity outside the ecliptic plane, producing a nearly polar two-body orbit with $i=89.4\degr$. The angle between $\mathbf{v}_{\infty,\oplus}$ and Earth's heliocentric velocity is $134.8\degr$; at fixed $\vinfearth$ and Earth speed, a bound solution at the escape threshold would require an angle of about $157.8\degr$, a rotation of roughly $23\degr$. This is much larger than the adopted $1.124\degr$ per-axis radiant perturbation, but it makes radiant fidelity central to the interpretation.

\begin{deluxetable}{lc}
\tablecaption{Nominal asymptotic directions and two-body heliocentric orbital elements for \polarim{}.\label{tab:orbit}}
\tablehead{\colhead{Quantity} & \colhead{Value}}
\startdata
Geocentric asymptotic velocity direction (RA, Dec) & $(139.1\degr,\,+59.5\degr)$ \\
Geocentric radiant (RA, Dec) & $(319.1\degr,\,-59.5\degr)$ \\
Heliocentric velocity direction (RA, Dec) & $(191.5\degr,\,+65.6\degr)$ \\
Heliocentric incoming direction (RA, Dec) & $(11.5\degr,\,-65.6\degr)$ \\
Heliocentric eccentricity $e$ & 1.92 \\
Perihelion distance $q$ & 0.91~au \\
Inclination $i$ & $89.4\degr$ \\
\enddata
\tablecomments{These values are nominal two-body descriptors of the pre-encounter heliocentric trajectory and are not a substitute for an $N$-body backward integration through the Earth--Moon system.}
\end{deluxetable}

\subsection{Monte-Carlo Robustness}

\begin{figure}
\centering
\includegraphics[width=\columnwidth]{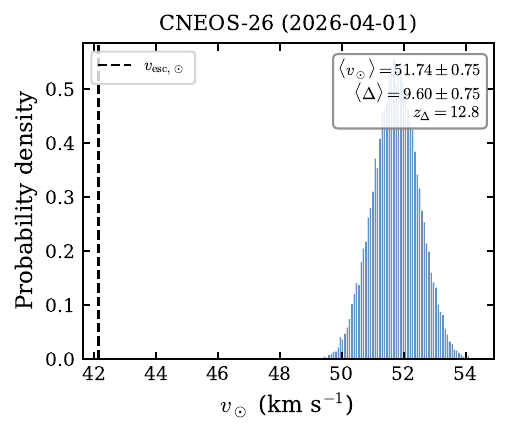}
\caption{Monte-Carlo distribution of heliocentric speed $\vhel$ for \polarim{}, based on realizations of the \citet{PenaAsensio2025} low-discrepancy uncertainty model. The dashed vertical line marks the solar escape speed $\vescsun$ at the event's heliocentric distance. The entire distribution lies above Solar System escape, with no bound realizations observed. The annotation reports the mean heliocentric speed, the mean margin above escape $\langle\Delta\rangle$, and the margin-to-scatter ratio $z_\Delta$.}
\label{fig:mc_hist}
\end{figure}

Using the propagation procedure defined in Section~\ref{sec:mcprop} and the low-discrepancy uncertainty model defined in Section~\ref{sec:errormodel}, the $N = 1{,}000{,}000$-realization Monte-Carlo run gives zero bound realizations ($k_\mathrm{bound} = 0$, $\hat{p}_\mathrm{iso} = 1.00000$). The corresponding 95\% finite-sample Monte-Carlo upper limit is $\pbound < 3.0\times10^{-6}$, with a lower confidence bound of $>99.9997\%$ on the unbound fraction, conditional on the adopted perturbation model. The Monte-Carlo heliocentric speed is $\langle\vhel\rangle = 51.74 \pm 0.75\kms$, the excess speed is $\langle\vinfsun\rangle = 30.00 \pm 1.29\kms$, and the margin is $\langle\Delta\rangle = 9.60 \pm 0.75\kms$, corresponding to a $12.82\sigma$ margin-to-scatter ratio. The continuous margin statistic also remains far from the bound/unbound boundary: the mean heliocentric speed is 12.82 Monte-Carlo standard deviations above escape under the adopted perturbation model. The Monte-Carlo heliocentric-speed distribution is shown in Figure~\ref{fig:mc_hist}.

\subsection{Statistical Interpretation}

The Monte-Carlo result $k_\mathrm{bound} = 0$ should be interpreted as a finite-sample upper limit on the bound probability, not as proof that the bound probability is exactly zero. For zero bound realizations in $N$ trials, the rule-of-three approximation gives $\pbound \lesssim 3/N$. With $N = 1{,}000{,}000$, this is $\pbound < 3.0\times10^{-6}$ at approximately 95\% confidence; the exact Clopper-Pearson upper limit is $2.9957\times10^{-6}$, numerically equivalent at the precision reported here. Equivalently, the model-conditional lower confidence bound on the unbound fraction is $>99.9997\%$. This limit is conditional on the adopted Gaussian low-discrepancy perturbation model.

The speed-margin statistic is independent of this binomial counting limit. With $\langle\Delta\rangle = 9.60 \pm 0.75\kms$, the adopted uncertainty scale would need to be inflated by a factor of roughly 12.8 before the mean margin would fall to $1\sigma$ in a first-order linear scaling. The direct uncertainty-inflation curve is the authoritative sensitivity calculation because the mean margin can drift under large angular perturbations.

\begin{deluxetable}{lc}
\tablecaption{Deterministic and Monte-Carlo results for \polarim{}.\label{tab:summary}}
\tablehead{\colhead{Property} & \colhead{Value}}
\startdata
\cutinhead{Deterministic}
Inertial geocentric speed $|\mathbf{v}_\oplus|$ (km\,s$^{-1}$) & 69.10 \\
Geocentric excess speed $\vinfearth$ (km\,s$^{-1}$) & 68.20 \\
$(v_x, v_y, v_z)_\odot$ (km\,s$^{-1}$) & $(-20.97,\,-4.28,\,+47.09)$ \\
Positive heliocentric $z$ component (km\,s$^{-1}$) & $+47.09$ \\
Heliocentric speed $\vhel$ (km\,s$^{-1}$) & 51.73 \\
Solar escape speed $\vescsun$ (km\,s$^{-1}$) & 42.14 \\
Speed margin $\Delta$ (km\,s$^{-1}$) & 9.59 \\
$\varepsilon_\odot$ (km$^2$\,s$^{-2}$) & $+450.1$ \\
$\vinfsun$ (km\,s$^{-1}$) & 30.00 \\
\cutinhead{Monte Carlo ($N = 1{,}000{,}000$)}
Bound draws $k_\mathrm{bound}$ & 0 \\
Unbound fraction $\hat{p}_\mathrm{iso}$ & 1.00000 \\
$\pbound$ (95\% finite-sample MC UL) & $< 3.0\times10^{-6}$ \\
Unbound fraction (95\% finite-sample MC LL) & $>99.9997\%$ \\
$\langle\vhel\rangle \pm \sigma_{\vhel}$ (km\,s$^{-1}$) & $51.74 \pm 0.75$ \\
$\langle\vinfsun\rangle \pm \sigma$ (km\,s$^{-1}$) & $30.00 \pm 1.29$ \\
$\langle\Delta\rangle \pm \sigma_\Delta$ (km\,s$^{-1}$) & $9.60 \pm 0.75$ \\
$z_\Delta$ margin-to-scatter ratio & 12.82 \\
\enddata
\end{deluxetable}

\section{Discussion}
\label{sec:discussion}

\subsection{Strength of the Candidate}

\polarim{} is a strong polar interstellar meteor candidate under the adopted low-discrepancy CNEOS uncertainty model. Its nominal heliocentric speed exceeds the solar escape speed at Earth's heliocentric distance by 9.59~km\,s$^{-1}$, and the Monte-Carlo margin-to-scatter ratio is $z_\Delta = 12.82$, i.e., a $12.82\sigma$ excess above escape under the adopted perturbation model. This is a larger post-2018 margin than obtained in the earlier automated screening of the 2022-07-28 and 2025-02-12 candidates, which had reported Monte-Carlo margins of $5.18 \pm 0.60\kms$ and $3.22 \pm 0.58\kms$, respectively \citep{CloeteLoeb2026}.

The heliocentric excess speed $\vinfsun = 30.00\kms$ is physically plausible for an object drawn from the local interstellar population. It lies within the range spanned by the telescopically discovered interstellar objects 1I/\okina Oumuamua ($\vinfsun \approx 26\kms$; \citealt{Meech2017}), 2I/Borisov ($\vinfsun \approx 32\kms$; \citealt{Guzik2020}), and 3I/ATLAS ($\vinfsun \sim 58\kms$; \citealt{seligman2025discovery}), while being inferred from an atmospheric-entry event rather than a telescopic orbit. The known telescopic sample remains too small for strong population inference, but the \polarim{} value is not anomalous in that context. The main statistical caveat is that the reported $\pbound$ is a finite-sample upper limit under a model, not an absolute probability: zero bound draws out of 1{,}000{,}000 implies $\pbound < 3.0\times10^{-6}$ and an unbound-fraction lower confidence bound of $>99.9997\%$ at 95\% confidence only if the low-discrepancy Gaussian perturbation model is representative of the event's true measurement error.

\subsection{Atmospheric Deceleration}

CNEOS velocities are reported for the observed fireball, typically close to peak brightness rather than at the top of the atmosphere. The measured speed may therefore be lower than the pre-atmospheric entry speed because of drag and ablation. For interstellar classification this bias is conservative: correcting for atmospheric deceleration would increase the inferred incoming speed and strengthen the unbound classification. \polarim{} was reported at 90.5~km altitude, higher than the altitudes of many large bolides. At this height the object may have experienced less deceleration before the reported measurement than lower-altitude events, but the available CNEOS metadata are not sufficient to model the deceleration history, and no atmospheric correction is applied.

\subsection{Velocity-Frame Assumption}

The calculation treats the CNEOS velocity components as Earth-fixed Cartesian components. This convention is supported by the public CNEOS fireball-table description, which defines the velocity-component frame as geocentric Earth-fixed with axes tied to Earth's rotation axis and prime meridian \citep{CNEOSFireballs}, and by the same ECEF convention used in the \citet{PenaAsensio2025} calibration. The SSD Fireball API field table itself is less explicit, labeling the components as Earth-centered entry-velocity components \citep{CNEOSFireballAPI}, so the calculation remains contingent on the public table description being the correct frame convention for the API values. Under the adopted Earth-fixed convention, the residual uncertainty associated with Earth-orientation corrections is far below the km\,s$^{-1}$ CNEOS velocity uncertainty and far below the candidate's 9.6~km\,s$^{-1}$ margin above escape. A wholesale inertial-versus-Earth-fixed axis misinterpretation would be a different systematic error and would change the heliocentric vector calculation; we do not treat that possibility as a small rotation correction. Confirmation of the exact provider convention and event-specific processing would further strengthen the interpretation.

\subsection{Error-Model Dependence}

\begin{figure}
\centering
\includegraphics[width=\columnwidth]{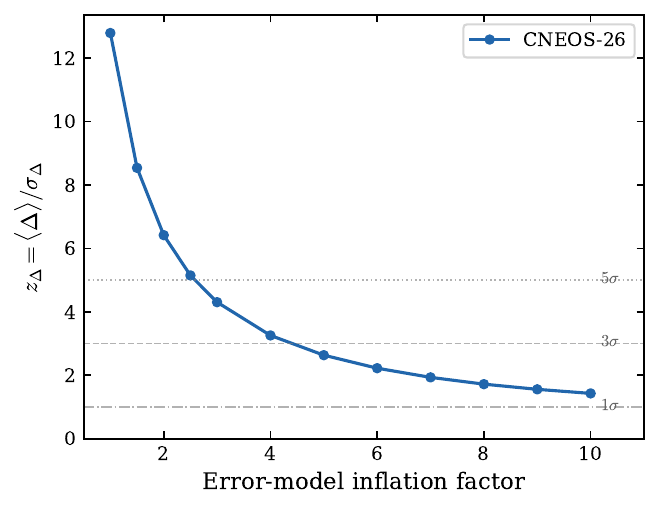}
\caption{Sensitivity of the interstellar classification to systematic underestimation of CNEOS velocity uncertainties. The margin-to-scatter ratio $z_\Delta$ is plotted as a function of a multiplicative inflation factor applied uniformly to all three uncertainty components ($\sigma_v$, $\sigma_\mathrm{RA}$, $\sigma_\mathrm{Dec}$). \polarim{} maintains a high margin under substantial error inflation, consistent with its nominal $z_\Delta = 12.82$.}
\label{fig:robustness}
\end{figure}

All robustness claims are conditional on the \citet{PenaAsensio2025} low-discrepancy model. \polarim{} qualifies for that regime because it occurred well after the 2018 transition used by the calibration. The adopted one-sigma speed error is 0.55~km\,s$^{-1}$, and the effective radiant-direction perturbation is about $1.12\degr$ per tangent-plane axis. Applying this calibration to a 2026 event assumes that the sensor and processing performance that defined the post-2018 low-discrepancy regime remained applicable through the event epoch. The large speed margin makes this candidate comparatively insensitive to modest error inflation. Since $z_\Delta = 12.82$, the uncertainty scale would have to be inflated by a factor of roughly 12.8 for the mean margin to fall to $1\sigma$ in a first-order linear scaling; a factor of four inflation would still leave the mean margin above $3\sigma$ in that approximation. The direct uncertainty-inflation curve is the authoritative sensitivity result because the mean margin can drift under large angular perturbations (Figure~\ref{fig:robustness}). The present analysis uses 1{,}000{,}000 Monte-Carlo draws, sufficient to establish a model-conditional finite-sample 95\% upper limit of $3.0\times10^{-6}$ on the bound fraction and a corresponding lower confidence bound of $>99.9997\%$ on the unbound fraction. Targeted uncertainty-inflation and importance-sampled tail calculations would more directly answer the physical question of how large a speed or radiant error is required to make the orbit bound.

The dominant remaining risk is not a small Gaussian perturbation, but a gross, event-specific, non-Gaussian error in the reported velocity vector. \citet{PenaAsensio2025} emphasize that hyperbolic CNEOS candidates should be interpreted with caution because velocities and inclinations in that subset may reflect measurement errors, and the CNEOS14 reliability debate illustrates why individual hyperbolic candidates require careful statistical treatment \citep{SocasNavarro2024}. \polarim{} has the second-largest geocentric speed in the analyzed 356-event sample and is therefore an outlier in the catalog context. The Monte-Carlo analysis demonstrates robustness to the calibrated low-discrepancy perturbations; it does not independently rule out a catastrophic velocity-vector failure.

\subsection{Recovery and Follow-Up}

The reported coordinates place the event over the South Atlantic east of Argentina. The impact energy is modest (0.086~kt) and the reported altitude is high (90.5~km). Those two facts make material recovery less straightforward than for a larger, lower-altitude bolide such as IM1 \citep{Loeb2024}. The event may have fragmented high in the atmosphere, and any surviving material would require a fall-ellipse calculation before the feasibility of a search could be assessed.

The first follow-up priority is therefore a higher-fidelity reconstruction rather than an expedition: (1)~produce targeted uncertainty-inflation and tail-sampling tests for the speed and radiant errors required to cross the bound threshold; (2)~back-integrate the trajectory with an $N$-body Earth-Moon-Sun model; (3)~refine the inbound radiant and asymptotic heliocentric velocity vector in the same dynamical model; and (4)~if warranted, model atmospheric entry, fragmentation, and wind drift to estimate a fall footprint. Independent validation is especially important because it is the main way to test the gross-outlier failure mode: any ground-based optical, infrasound, seismic, satellite, or regional fireball-network observation near 2026-04-01 02:13:14~UTC could help test the CNEOS velocity and radiant, and a provider-supplied per-event covariance or processing note would be the most useful single improvement.

\subsection{Implications}

If confirmed, \polarim{} would add a high-margin, nearly polar, post-2018 interstellar meteor to the CNEOS record. Its $\vinfsun$ of about 30~km\,s$^{-1}$ lies in the same broad range as known macroscopic interstellar objects and is high enough to make the unbound classification robust under the adopted small-error model. The event also demonstrates the value of continuous automated screening: the analysis workflow can ingest new CNEOS entries, compute heliocentric energy, apply the calibrated uncertainty model, and identify candidates rapidly enough that follow-up data searches can begin while auxiliary observations may still be available. That operational value is separate from any single event's final status---even candidates that later fail under improved calibration help define what information is needed for rigorous interstellar-meteor confirmation.

\section{Summary and Conclusions}
\label{sec:summary}

We report \polarim{}, the CNEOS fireball detected on 2026-04-01 02:13:14~UTC, as a high-likelihood polar interstellar meteor candidate under an automated heliocentric-energy analysis and the adopted post-2018 low-discrepancy uncertainty model of \citet{PenaAsensio2025}. The main findings are:

\begin{enumerate}
\item The event occurred at $-41.9\degr$, $-54.7\degr$, at a reported altitude of 90.5~km over the South Atlantic east of Argentina. Its reported Earth-fixed velocity vector is $(+3.6,\,-34.6,\,+59.8)\kms$, with a dominant positive $z$ component.
\item The deterministic heliocentric calculation gives $\vhel = 51.73\kms$, compared with a solar escape speed of $42.14\kms$ at Earth's heliocentric distance. The heliocentric J2000 $z$ component is $+47.09\kms$, already exceeding the scalar local escape speed, and the two-body heliocentric inclination is $89.4\degr$. The implied heliocentric specific energy is $+450.1\kmssq$, and the heliocentric excess speed is $\vinfsun = 30.00\kms$.
\item Under 1{,}000{,}000 Monte-Carlo realizations using the low-discrepancy CNEOS uncertainty model, no realization becomes bound to the Sun, giving a 95\% finite-sample Monte-Carlo upper limit $\pbound < 3.0\times10^{-6}$ and an equivalent lower confidence bound of $>99.9997\%$ on the unbound fraction, conditional on the adopted Gaussian perturbation model.
\item The Monte-Carlo speed margin above escape is $\langle\Delta\rangle = 9.60 \pm 0.75\kms$, giving a $12.82\sigma$ margin-to-scatter ratio under the adopted perturbation model. This makes \polarim{} robust against moderate underestimation of the calibrated velocity and radiant uncertainties.
\end{enumerate}

These conclusions are conditional on three key assumptions: the CNEOS velocity components follow the public Earth-fixed frame definition, the post-2018 low-discrepancy uncertainty model applies to this 2026 event, and the two-body Earth and Sun approximations are adequate for first-order classification. The Monte-Carlo result strongly disfavors a bound orbit under the calibrated small-error model, but it does not by itself rule out a rare event-specific, non-Gaussian velocity-vector failure. That gross-outlier risk remains the principal target for independent validation. Recommended next steps are to produce targeted uncertainty-inflation and tail-sampling tests, perform $N$-body backward integration through the Earth-Moon system, and search for independent observations near the event time. If the trajectory remains robust after those checks, \polarim{} should be treated as one of the strongest CNEOS interstellar meteor candidates in the calibrated post-2018 era.

\begin{acknowledgments}
This work was supported by the Galileo Project at Harvard University. The reported calculations made use of Astropy \citep{Astropy2022}, the JPL Horizons system \citep{Giorgini1996}, NumPy \citep{numpy2020}, and SciPy \citep{scipy2020}.
\end{acknowledgments}

\software{
Astropy \citep{Astropy2022},
NumPy \citep{numpy2020},
SciPy \citep{scipy2020},
Astroquery \citep{Ginsburg2019}
}

\bibliography{references}{}
\bibliographystyle{aasjournalv7}

\end{document}